\documentclass[aps,prl,twocolumn,superscriptaddress,showpacs,floatfix]{revtex4-1}
\usepackage{color}
\usepackage{dcolumn}   
\usepackage{bm}        
\usepackage{amssymb}   
\usepackage{amsmath}
\usepackage{gensymb}   
\usepackage{bbm}
\usepackage{hyperref}
\usepackage{graphicx}
\usepackage{array}
\usepackage{braket}
\usepackage{multirow}
\usepackage{natbib}
\newcommand{\bea}{\begin{eqnarray}}
\newcommand{\eea}{\end{eqnarray}}
\usepackage{overpic}

\begin{document}

\title{\texorpdfstring{Octupolar vs N\'{e}el Order in Cubic 5\textit{d}\textsuperscript{2} double perovskites}{Octupolar vs Neel Order in Cubic 5d2 double perovskites}}

\author{D. D. Maharaj}
\email{maharadd@mcmaster.ca}
\affiliation{Department of Physics and Astronomy, McMaster University, Hamilton, ON L8S 4M1 Canada}

\author{G. Sala}
\affiliation{Department of Physics and Astronomy, McMaster University, Hamilton, ON L8S 4M1 Canada}
\affiliation{Neutron Scattering Division, Oak Ridge National Laboratory, Oak Ridge, Tennessee 37831, USA}

\author{M. B. Stone}
\affiliation{Neutron Scattering Division, Oak Ridge National Laboratory, Oak Ridge, Tennessee 37831, USA}

\author{E. Kermarrec}
\affiliation{Department of Physics and Astronomy, McMaster University, Hamilton, ON L8S 4M1 Canada}
\affiliation{Laboratoire de Physique des Solides, CNRS, Univ.  Paris-Sud, Universit\'{e} Paris-Saclay, 91405 Orsay Cedex, France}

\author{C. Ritter}
\affiliation{Institut Laue-Langevin, Bo\^{i}te Postale 156, 38042 Grenoble C\'{e}dex, France}

\author{F. Fauth}
\affiliation{CELLS-ALBA Synchrotron, Carrer de la Llum 2-26, 08290 Cerdanyola del Vall\`{e}s, Barcelona, Spain}

\author{C. A. Marjerrison}
\affiliation{Brockhouse Institute for Materials Research, McMaster University, Hamilton, ON L8S 4M1 Canada}

\author{J. E. Greedan}
\affiliation{Brockhouse Institute for Materials Research, McMaster University, Hamilton, ON L8S 4M1 Canada}
\affiliation{Department of Chemistry and Chemical Biology, McMaster University, ON, L8S 4M1, Canada}

\author{A. Paramekanti}
\affiliation{Department of Physics, University of Toronto, 60 St. George Street, Toronto, ON, M5S 1A7 Canada}

\author{B. D. Gaulin}
\affiliation{Department of Physics and Astronomy, McMaster University, Hamilton, ON L8S 4M1 Canada}
\affiliation{Brockhouse Institute for Materials Research, McMaster University, Hamilton, ON L8S 4M1 Canada}
\affiliation{Canadian Institute for Advanced Research, 661 University Ave., Toronto, ON M5G 1M1 Canada}

\date{\today}

\begin{abstract}
We report time-of-flight neutron spectroscopic and diffraction studies of the 5$d^2$ cubic double pervoskite magnets, Ba$_2$\textit{M}OsO$_6$ (M = Zn, Mg, Ca).  These cubic materials are all described by antiferromagnetically-coupled $5d^2$ Os$^{6+}$ ions decorating a face-centred cubic (FCC) lattice.  They all exhibit thermodynamic anomalies consistent with phase transitions at a temperature $T^*$, and exhibit a gapped magnetic excitation spectrum with spectral weight concentrated at wavevectors typical of type I antiferromagnetic orders.  
While muon spin resonance experiments show clear evidence for time reversal symmetry breaking, no corresponding magnetic Bragg scattering is observed at low temperatures. 
These results are argued to be consistent with low temperature octupolar order, and are discussed in the context of other $5d$ DP magnets, and theories for $d^2$ ions on a FCC 
lattice which predict exotic orders driven by multipolar interactions.
\end{abstract}

\pacs{75.25.−j, 75.40.Gb, 75.70.Tj}

\maketitle

\textit{Introduction} --- Ordered double perovskite (DP) magnets, with the chemical formula $A_2BB^{\prime}O_6$, 
provide a fascinating avenue for the study of geometric frustration and its interplay with strong spin-orbit coupling
(SOC) \cite{Krempa2014}. Here, $B$ and $B^{\prime}$ sublattices individually form an FCC lattice of edge-sharing tetrahedra, 
an archetype for geometric frustration in three dimensions. Furthermore, the flexibility of the DP lattice to host 
heavy ions at the $B^{\prime}$ site allows the study of spin-orbit driven physics, as the strength of SOC scales 
$\sim Z^2$, where $Z$ is the atomic number of the magnetic ion. This interplay of SOC and frustration has been predicted
to yield exotic ground states in a host of these DP systems 
\cite{Krempa2014, ChenBalents2010, ChenBalents2011, Svoboda_PRB2017, Svoboda_arXiv2017}.  

Strong SOC splits the $t_{2g}$ levels associated with the magnetic $B^{\prime}$ in an octahedral crystal field, resulting in a four-fold degenerate, $j \!=\! 3/2$, ground state and a doubly degenerate, $j \!=\! 1/2$, excited state.  Famously, for a $d^5$ electronic configuration, as occurs for Ir$^{4+}$ or Ru$^{3+}$, this results in a single hole in a $j \!=\! 1/2$ state, leading to extreme quantum magnetism, and bond-dependent Kitaev interactions in appropriate geometries \cite{JackeliPRL2010,Singh2012,Plumb2014,BJKim_NPhys2015,BanerjeeRuCl3_2016,Winter_2017,Kasahara2018}.  

In the case of $d^2$ ions on the $B^{\prime}$ site, the subject of this Letter, the combination of SOC and Hund's coupling results in an effective $J \!=\! 2$ angular momentum. Studies of such interacting $d^2$ ions on the FCC lattice have highlighted the importance of orbital repulsion, 
in addition to the conventional magnetic exchange. This leads to complex multipolar interactions, and large regimes of quadrupole order in the 
predicted phase diagram \cite{ChenBalents2010,ChenBalents2011,Svoboda_PRB2017,Svoboda_arXiv2017}.

Here we report new magnetic neutron powder diffraction (NPD), inelastic neutron scattering (INS), and high resolution X-ray diffraction (XRD) results 
on three cubic DPs with a 5$d^2$ electronic configuration at the $B^{\prime}$ site: Ba$_2$\textit{M}OsO$_6$, with $M$ = Zn, Mg, Ca 
(respectively referred to henceforth as BZO, BMO, and BCO).  All three display clear thermodynamic signatures of a phase transition
\cite{Thompson_JPCM2014,Kermarrec2015,MarjerrisonPRB2016} 
at relatively high temperatures $\sim\! 30$-$50$\,K, which is associated with time-reversal symmetry breaking based on oscillations observed in zero
field muon spin relaxation ($\mu$SR) experiments \cite{MarjerrisonPRB2016}.  Our INS results show strong, gapped, magnetic spectral weight at wavevectors typical of type I antiferromagnetic (AF) order, but we find no indication of an ordered moment in the diffraction data, allowing us to place upper limits of between 
$0.13$-$0.06$ $\mu_B$ per $B^{\prime}$ site. Furthermore, our NPD and XRD results show no deviation from cubic symmetry expected from
quadrupolar ordering. We present arguments showing that these results may be understood as arising from 
octupolar order in these cubic 5$d^2$ DP materials.

Multipolar ordered phases have been extensively studied in $f$-electron compounds 
\cite{MultipolarRMP2009}. Examples include NpO$_2$ \cite{SantiniNpO2_PRL2000, NpO2TripleQ_PRL2002, Fazekas_PRB2003, NpO2NMR_PRL2006}, where experiments have been interpreted in terms of a primary rank-$5$ magnetic multipolar order
(triakontadipoles) driving secondary quadrupolar order, the ``hidden order" 
state of URu$_2$Si$_2$ which has been proposed to host hexadecapolar order
\cite{HauleKotliar2009}, and recent discoveries of quadrupolar and octupolar orders in PrX$_2$Al$_{20}$ (X = Ti, V)
\cite{Sakai_JPSJ2011,Sato_PRB2012,Nakatsuji_PRL2014}.
In stark contrast, multipolar orders in $d$-electron systems are rare 
\cite{LFu_PRL2015, Hsieh_Science2017, Mitrovic_NComm2017, Motome2018, Mitrovic_Physica2018, Hiroi_JPSJ2019}; our work appears to be the
first report of octupolar order.

\begin{figure}[t]
\centering
\includegraphics[width=8.5cm]{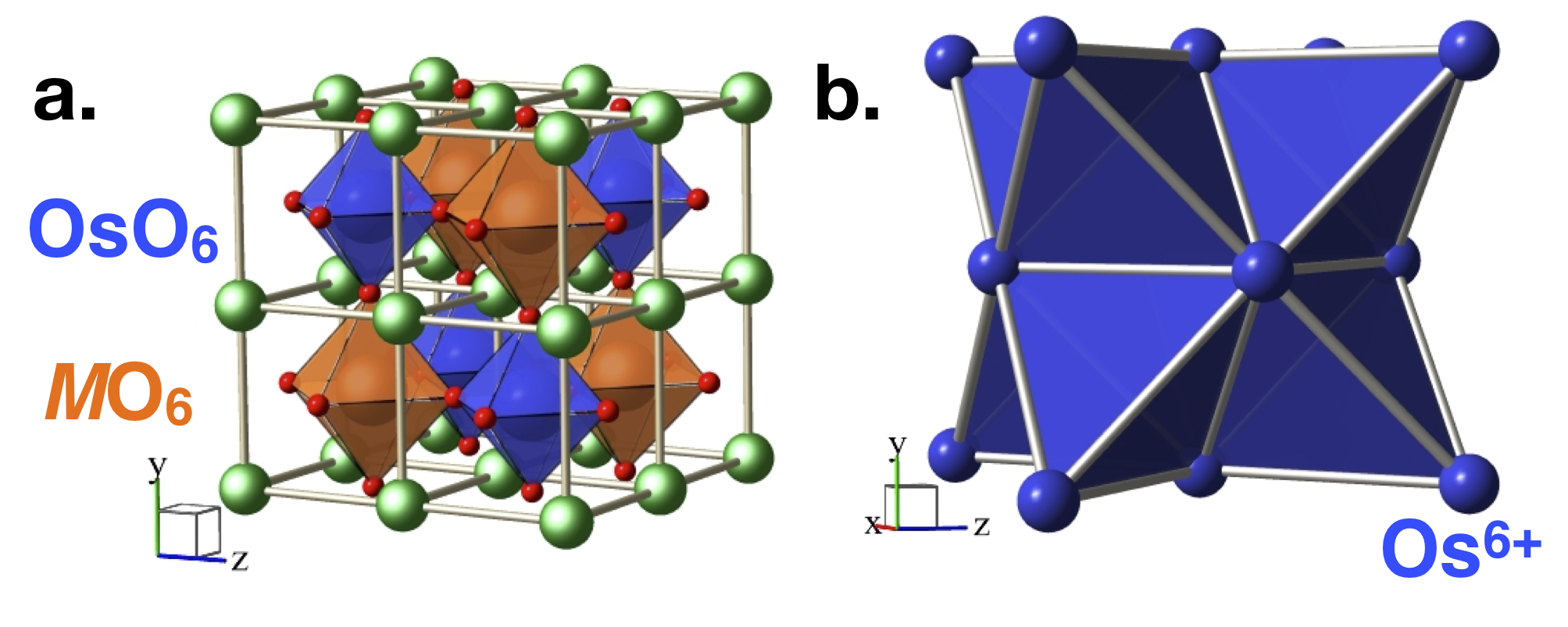}%
\caption{(a) Structure of cubic double perovskites with space group $Fm\bar{3}m$. (b) The frustrated FCC network of edge-sharing tetrahedra of the B$^\prime$ site magnetic moments that are generated in the cubic  Ba$_2$BB$^{\prime}$O$_6$ structure.}
\label{fig:1} 
\end{figure}

\begin{table}[b]
\begin{ruledtabular}
\begin{tabular}{cccccc}
System & $T^*$ & $\theta_{CW}$ & \textit{a} (\AA) & Ref. & $\mu_{ord}$ \\
\hline
$\mathrm{Ba_2CaOsO_6}$ & 49  & --$156.2(3)$ & 8.3456 & [13] & $<$ 0.11$\mu_B$ \\
\hline
$\mathrm{Ba_2MgOsO_6}$ & 51 & --$120(1)$ & 8.0586 & [15] & $<$ 0.13$\mu_B$ \\
\hline
$\mathrm{Ba_2ZnOsO_6}$ & 30 & --$149.0(4)$ & 8.0786 & [15] & $<$ 0.06$\mu_B$ \\
\end{tabular}
\end{ruledtabular}
\caption{Summary of experimental results for the three cubic DPs studied. $T^*$ denotes the peak in the heat capacity indicating a thermodynamic phase transition 
\cite{Thompson_JPCM2014,MarjerrisonPRB2016}. $\theta_{CW}$ is the Curie-Weiss temperature extracted from high temperature susceptibility data 
\cite{Thompson_JPCM2014,MarjerrisonPRB2016}. $\mu_{ord}$ is the upper limit on the ordered dipolar moment associated with type-I AF order, as determined 
from neutron diffraction in this work.}
\label{table}
\end{table}

BZO, BMO and BCO have been previously studied in powder form.  In all three materials, neutron and 
x-ray diffraction confirm that they remain in the cubic $Fm\bar{3}m$ space group down to the lowest temperature.
They all display Curie-Weiss-like magnetic susceptibilities ($\chi$) at high temperatures, with large AF Curie-Weiss 
constants ($\Theta_{\rm CW} \! \sim \! 130$\,K), and anomalies
at $T^*$ in the form of a splitting between field-cooled and zero-field cooled results.  
They all exhibit peaks in their heat capacity, or in
the related measure 
$d(\chi T)/dT$, at $T^* \!\!\sim\!\! 50$\,K (BMO, BCO) or $T^* \!\! \sim \!\! 30$\,K (BZO), indicating a phase transition \cite{MarjerrisonPRB2016,Thompson_JPCM2014}.
These findings are summarized in Table I.

The entropy released up to $\sim \! 2 T^*$ in all three materials appears to be $\sim\! R \ln(2)$ per mole, much smaller than $R \ln (5)$ 
expected for a $J\!=\!2$ moment \cite{Thompson_JPCM2014,Kermarrec2015,MarjerrisonPRB2016}.
This points to part of the entropy being quenched at $T \gg T^*$.
This is in contrast to the $\sim\! R \ln (5)$ entropy released up to $\sim 2 T_N$ for the tetragonal counterpart Sr$_2$MgOsO$_6$, which has a 
high N\'eel ordering temperature $T_N \sim 100$\ K \cite{Morrow_SciRep2016}.

These three cubic samples have also been previously studied using $\mu{SR}$ techniques \cite{Thompson_JPCM2014,MarjerrisonPRB2016}, and it is primarily on the basis of these 
zero longitudinal field (ZLF) $\mu {SR}$ oscillations for $T < T^*$, indicative of a time-reversal broken state, that the transition at $T^*$ was associated with AF order.  However, no magnetic 
neutron diffraction peaks could be identified in this earlier study at low temperatures, with a sensitivity to ordered moment of $\sim$ 0.7 $\mu_B$. In the present work, we significantly
improve on this bound, still finding no evidence of magnetic Bragg peaks.

The corresponding 5$d^3$ osmium-based DP magnets, both cubic Ba$_2$YOsO$_6$ and monoclinic Sr$_2$ScOsO$_6$ and La$_2$LiOsO$_6$, show clear N\'{e}el transitions to AF ordered states, with large ordered moments of $\sim\! 1.7 \mu_B$ \cite{Maharaj2018,Kermarrec2015,Taylor_PRB2015,Thompson_PRB2016,Aczel_PRB2013}.  
These observed ordered moments are reduced from the  
$3 \mu_B$ value expected for an orbitally-quenched moment, pointing to strong SOC effects, or covalency, or both.  Nonetheless magnetic Bragg scattering at the (100) and (110) positions is easily observed, along with strong, gapped inelastic magnetic scattering centred at these two ordering wavevectors.  Previously studied 5$d^2$ DPs such as 
monoclinic Sr$_2$MgOsO$_6$ and cubic Ba$_2$LuReO$_6$ (with Re$^{5+}$) also show transitions to Type I AF order, as seen via neutron diffraction, albeit with much smaller 
ordered moments, $0.6(2)$ and $0.34(4)$ $\mu_B$, respectively \cite{Morrow_SciRep2016,Xiong_JSSC2018}.

Below we present our experimental findings on powder samples of the cubic systems, BZO, BMO and BCO. Details of experimental and analysis methods can be found in the Supplemental Material (SM) \cite{SM}. Our new NPD measurements on D20 \cite{D20} at the Institut Laue Langevin have $\sim$ 10 to 20 times more sensitivity to magnetic Bragg scattering as compared with previous NPD measurements taken at the C2 instrument of the Chalk River Laboratories.  No magnetic Bragg scattering is observed at $10$\,K, factors of $3$--$5$ below $T^*$ for any of these materials. We do however observe gapped, inelastic magnetic spectral weight centred on wavevectors characteristic of type I AF order, leading us to conclude that the dominant broken symmetry below $T^*$ 
in these three cubic DP $d^2$ magnets must involve multipolar ordered phases.

\noindent \textit{Results} --  Time-of-flight INS measurements which were performed at SEQUOIA \cite{Sequoia} are shown in Fig. 2. Panels (a)-(c) show the INS spectra well below (top panel) and above $T^*$ for BZO, BMO, and BCO respectively. Panels (d)-(f) show cuts through this data as a function of energy, integrating all $|Q| <$ 1.15 \AA$^{-1}$ and as a function of temperature, again for BZO, BMO and BCO respectively.

The data sets for all three samples in Fig. 2 are similar, with gapped magnetic spectral weight at low $|Q|$'s, typical of the 100 (0.78 \AA$^{-1}$) and 110 (1.1 \AA$^{-1}$) Bragg positions.  The full bandwidth of the magnetic excitation spectrum appears to be $\sim$ 6 meV.  From Fig. 2 b), c), e), and f), this magnetic spectral weight overlaps in energy 
with strong phonon scattering near $\sim$ 18 meV and 14 meV for BMO and BCO respectively.  Even though our low $|Q|$-integration favours magnetic scattering at the expense of scattering from phonons, whose intensity tends to go like $|Q|^2$, we still pick up a sizeable contribution from this high phonon density of states (DOS), especially at high temperatures where the thermal population of the phonons is large.  A redshift in the peak of the phonon DOS from $\sim$ 17 meV in BMO to $\sim$ 14 meV for BCO is observed. This is expected, as Ca$^{2+}$ is isoelectronic to Mg$^{2+}$ but heavier, hence, all other factors being equal, BCO will display lower frequency phonons. While the Zn$^{2+}$ in BZO is heavier still
than Ca$^{2+}$, it is not isoelectronic, instead possessing a filled 3$d$ shell.  This might lead to its higher energy phonon.

As the high phonon DOS is well separated from the magnetic spectral weight in BZO, shown in Figs.~2(a) and 2(d), this is where the nature of the gapped magnetic scattering can be most easily appreciated.  The energy cuts in Fig. 2(d) clearly show a well developed gap of $\sim$ 10 meV and a bandwidth of $\sim$ 6 meV.  This structure collapses by $25$\,K, where $T^*$ = 30 K for BZO, at which point the gap has largely filled in and only a vestige of an overdamped spin excitation at $\sim$ 10 meV remains.  This is very similar to what occurs in the $d^3$ DPs on the approach to their $T_N$s, {\it except} that there is no obvious temperature dependent Bragg scattering at the 100 or 110 positions, as 
would be expected for type I AF order.

\begin{figure}[htbp!]
\centering
\includegraphics[width=\columnwidth]{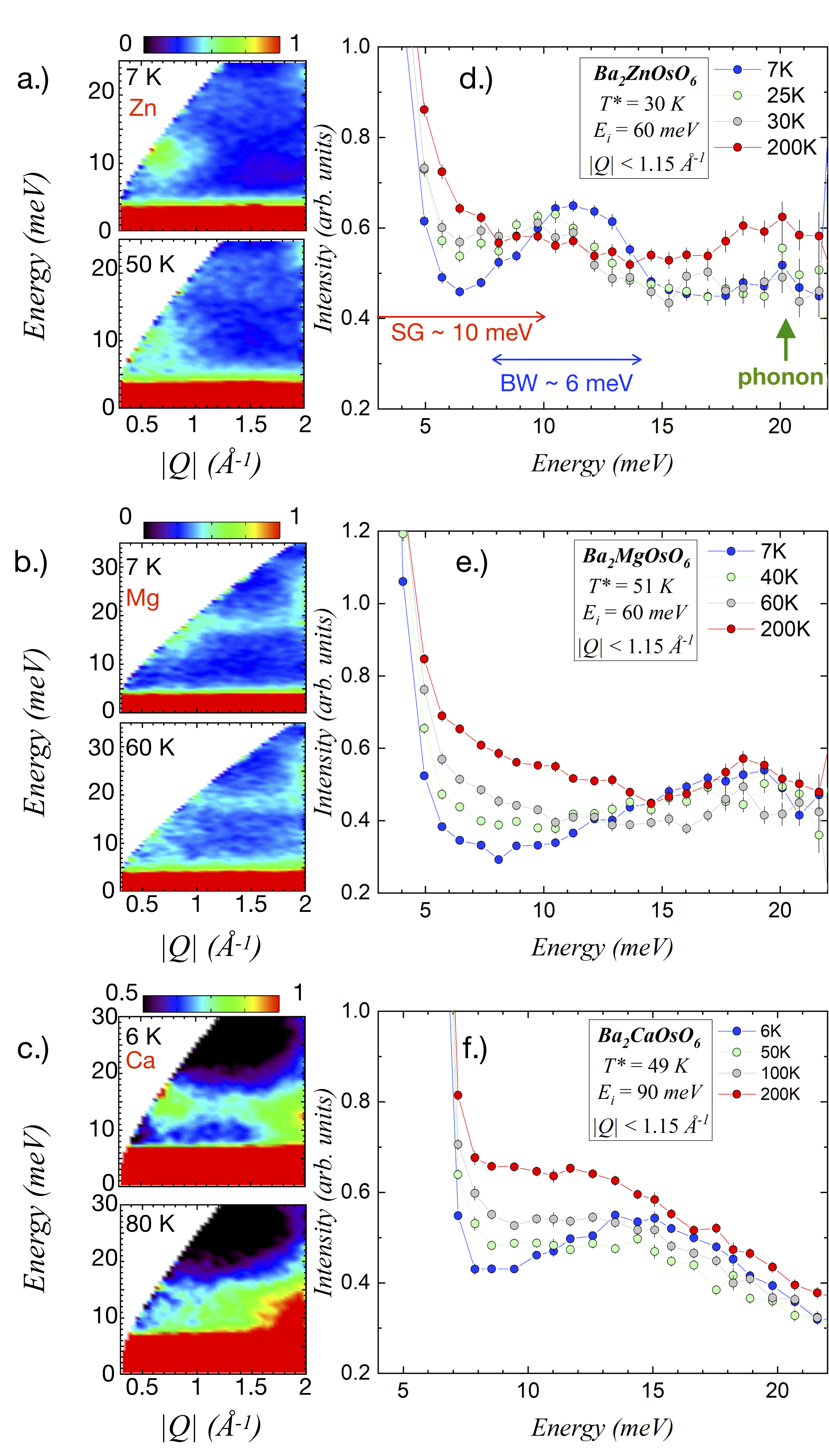}
\caption{\label{fig: INSData}(a) - (c): Neutron scattering intensity contour plots for BZO, BMO, and BCO shown as a function of energy transfer, \textit{E} and momentum transfer $|Q|$ at base temperature (top) and at $T > T^*$ (bottom), respectively. Below $T^*$, clear gapped magnetic inelastic spectral weight develops around (100) and (110) wavevectors ($\sim$ 0.78 \AA$^{-1}$) and 110 (1.1 \AA$^{-1}$) in each case. (d) and (e): Low $|Q|$ integrated cuts of the neutron scattering intensity as a function of energy transfer \textit{E} as a function of temperature for BZO, BMO, and BCO, respectively. A gap in the magnetic excitation spectum is clearly revealed for each compound for $T < T^*$.}
\end{figure}

The absence of evidence for magnetic Bragg scattering is seen in Fig.~3. Fig.~3(a) shows neutron diffraction data taken at T = 10 K, well below $T^*$ = 30 K in BZO, using the D20 diffractometer at the Institut Laue-Langevin \cite{D20}. This data and the corresponding NPD data on BMO and BCO refine in the cubic $Fm\bar{3}m$ space group at all temperatures measured.  Figure 3 b), c) and d) then show a subtraction of high temperature (50 K for BZO; 70 K for BMO and BCO) data sets from low temperature data sets for each of BZO, BCO, and BMO, respectively.  A calculated neutron diffraction profile appropriate for a type I AF structure below $T^*$ is shown as the red line in Fig. 4 b), c) and d), where the assumed ordered moment in the calculation is $0.06 \mu_B$ for BZO (b), $0.11  \mu_B$ for BMO (c), and $0.13  \mu_B$ for BCO (d).  Taking the case where the evidence {\it against} long range magnetic order below $T^*$ is most stringent, BZO, we can eliminate conventional type I AF order of magnetic dipoles with an ordered moment greater than $\! \sim\! 0.06 \mu_B$.  This upper limit for magnetic dipole order is a factor of $\sim\!12$ more stringent than previous limits on magnetic Bragg scattering for this family of 
cubic DP materials.  This upper bound for $\mu_{ord}$ in BCO is  $\sim\! 35\%$ lower than the value, $0.2 \mu_B$, which was previously extracted from a comparison of the 
$\mu{SR}$ internal fields of BCO and Ba$_2$YRuO$_6$ \cite{Thompson_JPCM2014,Carlo_PRB2013}.  

{\it Competing multipolar orders. ---} 
Our study shows all or most of the static 5$d^2$ moment associated with Os$^{6+}$ in BZO, BMO and BCO is not visible to neutron 
diffraction below the transition at $T^*$.  Nonetheless, strong inelastic magnetic scattering is easily observed at all temperatures, and 
it is most clearly gapped at low temperature.
One possible scenario to explain these results is that the ground state has dominant quadrupolar ordering, accompanied by weak dipolar ordering, as 
proposed in certain models \cite{ChenBalents2010,ChenBalents2011,Svoboda_arXiv2017}. 
In this scenario, the observed entropy is most naturally accounted for by assuming a two-step transition,
with quadrupolar ordering at $T \gg T^*$ partially quenching the $R \ln 5$ entropy, and the residual $\sim\!R \ln(2)$ entropy being quenched by AF dipolar ordering at $T^*$.
The quadrupolar order would favor occupation of a specific orbital, and pin the direction of the ordered dipole moment, resulting in time-reversal 
breaking and a spin-gap. If the ordered dipole moment is weak, it may escape detection in a NPD experiment. 
However, such a 
quadrupolar ordered state would lower the crystal symmetry due to orbital ordering, which is at odds with our high resolution NPD data shown for BCO in Fig.~4(a). We have carried out even higher-resolution synchrotron X-ray diffraction (XRD) measurements on BCO, the family member which best exhibits undamped ZF $\mu${SR} oscillations. The sensitivity of these measurements to possible weak splittings of the cubic Bragg peaks is $\sim\!10$ times 
greater than the NPD measurements, as the inset to Fig.~4(a) demonstrates. These XRD results, in Figs.~4 (b)-(d), show no splitting of the cubic Bragg peaks, confirming that BCO remains cubic even for $T \ll T^*$, inconsistent with quadrupolar ordering.

\begin{figure}[tb]
\centering
\includegraphics[width=\columnwidth]
                {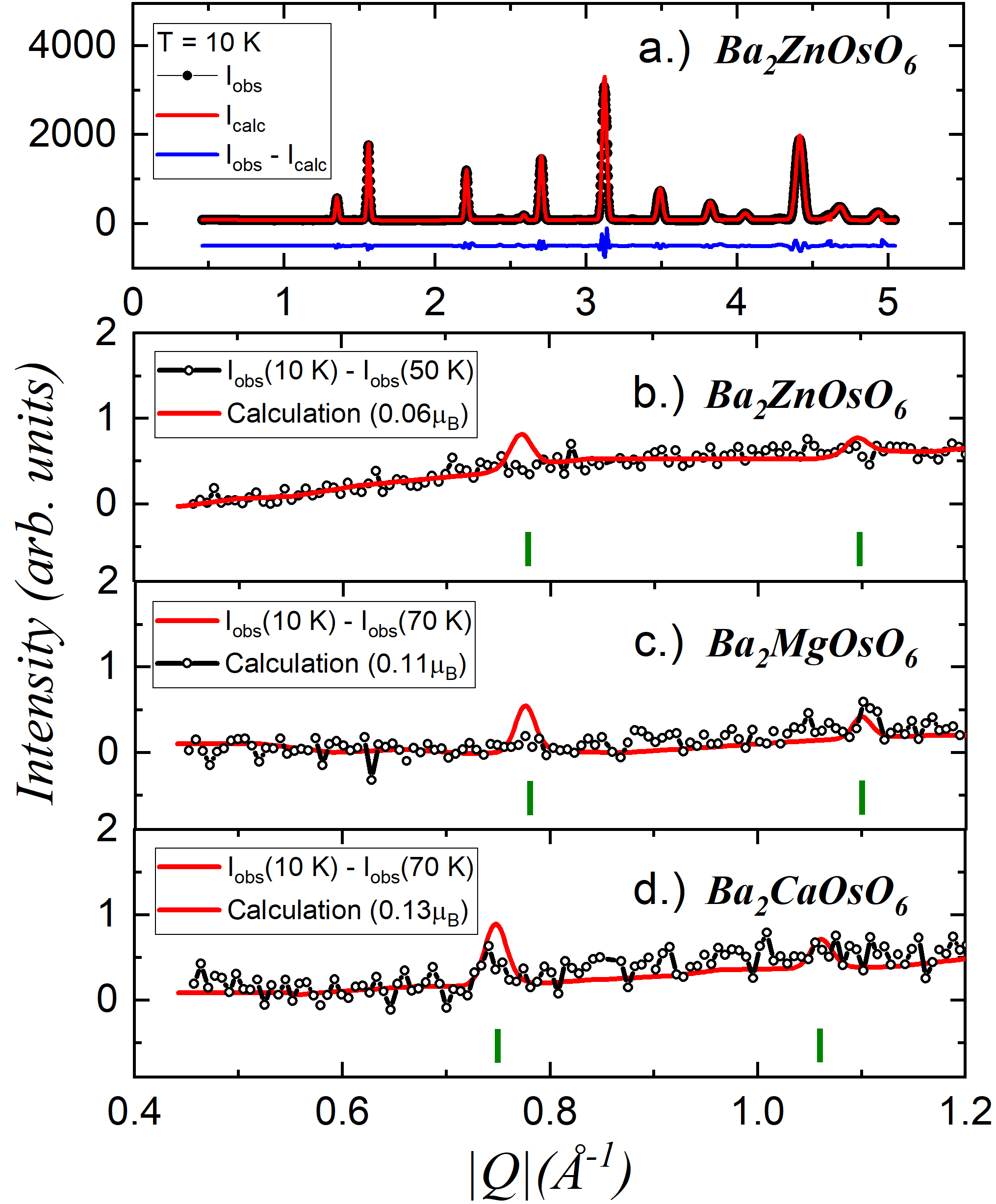}
\caption{\label{fig: ZnNPD}(a) NPD measurements on BZO for \textit{T} = 10 K with the experimental data set in black and the fit to the refined Fm$\bar{3}$m structure in red. (b) Subtraction of the 50 K data set from the 10 K data set. The red line shows the calculated magnetic diffraction pattern for BZO with an Os$^{6+}$ ordered moment of 0.06$\mu_B$, which we establish as the upper limit for an ordered dipole moment in BZO . Green fiducial lines indicate the locations of the magnetic peaks expected for type I AF order. Panels (c) and (d) show the same comparison for BMO and BCO. These establish upper limits on an ordered Os$^{6+}$ dipole moment of 0.11$\mu_B$ and 0.13$\mu_B$, respectively.}
\end{figure}

\begin{figure}[tb]
\includegraphics[width=\columnwidth]{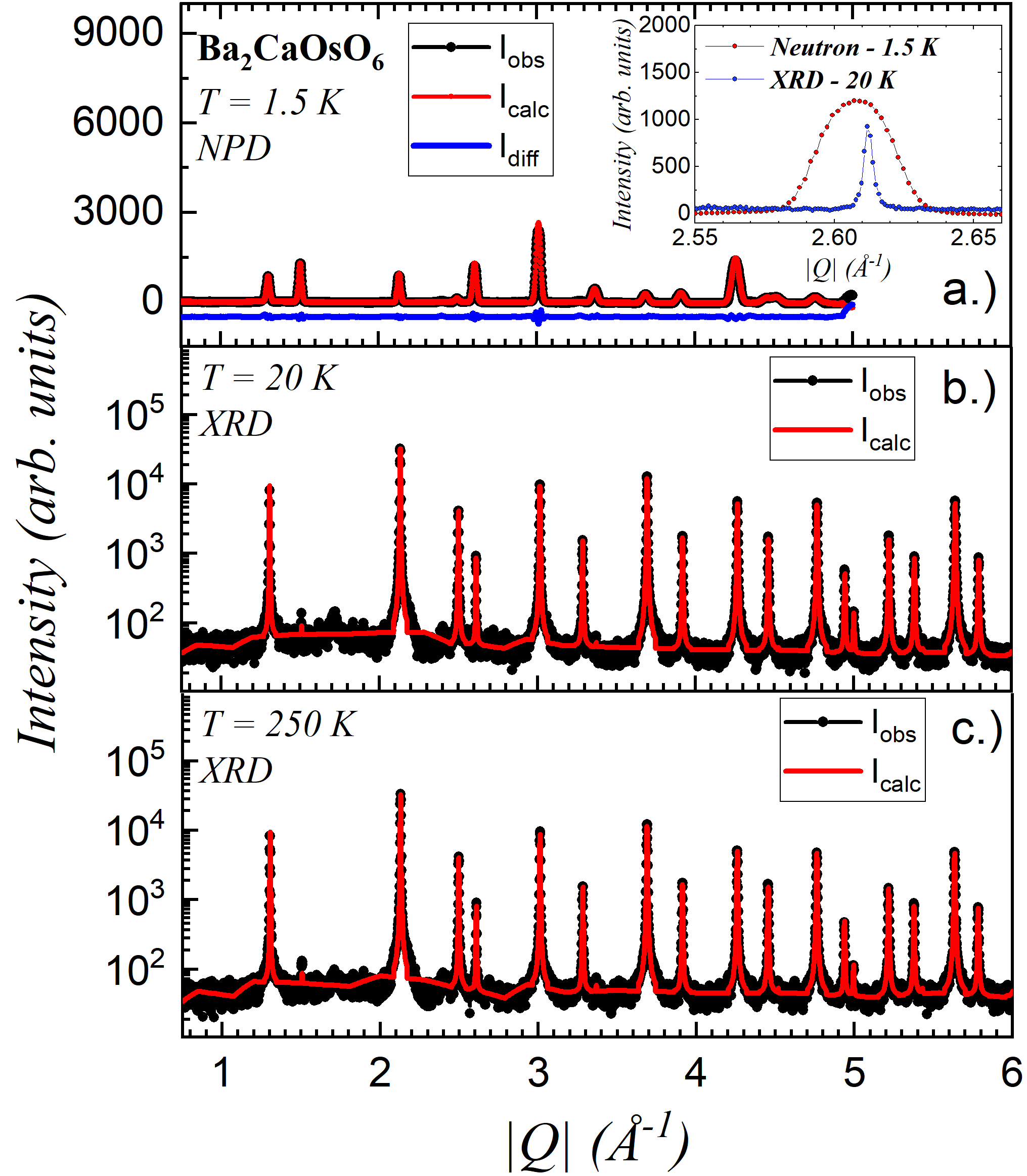}%
\caption{\label{fig: 5} a) The NPD profile for BCO is shown at $T$ = 1.5 K in the main panel, while the inset shows a comparison of neutron vs synchrotron x-ray diffraction data taken on BCO at 20 K.   Panels b) and c) show synchrotron x-ray scattering data on BCO at T = 20 K (b), and T = 250 K (c), along with corresponding cubic structural refinements, in red.}
\end{figure}

Here, we propose a distinct scenario - an octupolar ordered ground state - that provides the most promising vehicle to explain all the salient 
observations.
For an effective $J\!=\!2$ moment, a residual octahedral crystal field Hamiltonian is
$H_{\rm CEF} \!=\! - V_{\rm eff} ({\cal O}_{40} + 5  {\cal O}_{44})$,
where the Steven's operators (dropping constant terms) are
\bea
{\cal O}_{40} &=& 35 J^4_z - (30 J (J+1)-25) J_z^2 \\
{\cal O}_{44} &=& \frac{1}{2} (J_+^4  + J_-^4).
\eea
For $V_{\rm eff} > 0$, this results in a non-Kramers ground state doublet, and an excited
triplet with a gap $\Delta=120 V_{\rm eff}$. This naturally accounts for partial entropy quenching for $T \lesssim \Delta$, 
without a phase transition, with the residual $R\ln(2)$ entropy being quenched by ordering within the doublet sector at $T^*$.
In the $|J_z=m\rangle$ basis, the ground state wavefunctions  are
$|\psi_{g,\uparrow}\rangle = |0\rangle$ and $|\psi_{g,\downarrow}\rangle = \frac{1}{\sqrt{2}} (|2\rangle + | -2 \rangle)$,
with excited triplet wavefunctions $|\psi_{e,\pm}\rangle = |\pm 1\rangle$ and $|\psi_{e,0}\rangle =  \frac{1}{\sqrt{2}} (|2\rangle - | -2 \rangle)$.
The ground state manifold thus has vanishing matrix elements for the dipole operators $\vec J$, precluding magnetic dipole ordering. However,
$\vec J$ can induce transitions into the excited triplet, thus accounting for the spin gap seen in INS.
Defining pseudospin-$1/2$ operators $\vec\tau$ within the ground state doublet, the quadrupolar operators
$(J_x^2-J_y^2) \equiv 2\sqrt{3} \tau_x$, $(3 J_z^2 - J^2) \equiv - 6 \tau_z$, while the octupolar operator 
$\overline{J_x J_y J_z} \equiv - \sqrt{3}\tau_y$ (where the overline denotes symmetrization).
Thus, the ground state doublet can
lead to broken time-reversal symmetry below $T^*$, while preserving cubic symmetry,
if $\langle \tau_y \rangle \neq 0$. Further details are given in the SM \cite{SM} and in Ref.~\cite{ParamekantiPRB}.

To conclude, the low temperature phases of the cubic $5d^2$ DPs BZO, BMO, and BCO are best described as arising from a ground state 
non-Kramers doublet with octupolar symmetry breaking. This exotic ground state appears to require the perfect FCC structure as 
non-cubic $d^2$ DPs display more conventional AF ground state selection. Tools such as magnetostriction, which has recently been proposed
as a useful way to detect octupolar ordering \cite{Patri2019}, or other experiments, could be needed to provide a smoking gun signature of octupolar 
symmetry breaking in these $5d^2$ materials.

\section{Acknowledgments}
This work was supported by the Natural Sciences and Engineering Research Council of Canada.   It was also supported in part by the National Science Foundation under Grant No. PHYS-1066293 and the hospitality of the Aspen Center for Physics.  We also acknowledge the hospitality of the Telluride Science Research Center. We gratefully acknowledge useful conversations with G. M. Luke and G. Chen. We are very grateful for the instrument and sample environment support provided during our inelastic neutron scattering measurements at SEQUOIA. The experiments which were performed at the Spallation Neutron Source at Oak Ridge National Laboratory was sponsored by the US Department of Energy, Office of the Basic Energy Sciences, Scientific User Facilities Division. The authors would also like to acknowledge ILL for beam time allocation experiment code 5-31-2577, doi:10.5291/ILL-DATA.5-31-2577. We acknowledge the BL04-MSPD staff of ALBA for the x-ray synchrotron powder diffraction data collection

\bibliography{octupolar}

\end{document}


\title{Supplemental Material: \\
Octupolar vs N\'{e}el Order in Cubic $5 d^2$ Double Perovskites}
\author{D. D. Maharaj}
\email{maharadd@mcmaster.ca}
\affiliation{Department of Physics and Astronomy, McMaster University, Hamilton, ON L8S 4M1 Canada}

\author{G. Sala}
\affiliation{Department of Physics and Astronomy, McMaster University, Hamilton, ON L8S 4M1 Canada}
\affiliation{Neutron Scattering Division, Oak Ridge National Laboratory, Oak Ridge, Tennessee 37831, USA}

\author{M. B. Stone}
\affiliation{Neutron Scattering Division, Oak Ridge National Laboratory, Oak Ridge, Tennessee 37831, USA}

\author{E. Kermarrec}
\affiliation{Department of Physics and Astronomy, McMaster University, Hamilton, ON L8S 4M1 Canada}
\affiliation{Laboratoire de Physique des Solides, CNRS, Univ.  Paris-Sud, Universit\'{e} Paris-Saclay, 91405 Orsay Cedex, France}

\author{C. Ritter}
\affiliation{Institut Laue-Langevin, Bo\^{i}te Postale 156, 38042 Grenoble C\'{e}dex, France}

\author{F. Fauth}
\affiliation{CELLS-ALBA Synchrotron, Carrer de la Llum 2-26, 08290 Cerdanyola del Vall\`{e}s, Barcelona, Spain}

\author{C. A. Marjerrison}
\affiliation{Brockhouse Institute for Materials Research, McMaster University, Hamilton, ON L8S 4M1 Canada}

\author{J. E. Greedan}
\affiliation{Brockhouse Institute for Materials Research, McMaster University, Hamilton, ON L8S 4M1 Canada}
\affiliation{Department of Chemistry and Chemical Biology, McMaster University, ON, L8S 4M1, Canada}

\author{A. Paramekanti}
\affiliation{Department of Physics, University of Toronto, 60 St. George Street, Toronto, ON, M5S 1A7 Canada}

\author{B. D. Gaulin}
\affiliation{Department of Physics and Astronomy, McMaster University, Hamilton, ON L8S 4M1 Canada}
\affiliation{Brockhouse Institute for Materials Research, McMaster University, Hamilton, ON L8S 4M1 Canada}
\affiliation{Canadian Institute for Advanced Research, 661 University Ave., Toronto, ON M5G 1M1 Canada}

\date{\today}
\maketitle

Details relevant to both experimental and theoretical aspects of the main paper are discussed in this supplemental material. Experimental methods for the time-of-flight inelastic neutron scattering (INS), neutron powder diffraction (NPD) and high angular resolution x-ray powder diffraction (XRD) measurements are fully explained in section I., A, B and C. An outline of the calculations performed in order to estimate the upper bound on the ordered magnetic moment in Ba$_2$CaOsO$_6$, Ba$_2$MgOsO$_6$, and Ba$_2$ZnOsO$_6$ is provided in section I B. part 2. Section II provides further detail for the theoertical argumemnt which determines the ground and excited state wavefunctions for the $d^2$ ions in Ba$_2$CaOsO$_6$, Ba$_2$MgOsO$_6$, and Ba$_2$ZnOsO$_6$.

\section{Experiment Details}

\subsection{Time-of-flight Inelastic Neutron Scattering}

\noindent \textit{Methods} -- The INS measurements were conducted using the direct geometry chopper spectrometer SEQUOIA, which is located at the Spallation Neutron Source \cite{Sequoia} of Oak Ridge National Laboratory. Powder samples weighing 8 grams of Ba$_2$MgOsO$_6$, Ba$_2$ZnOsO$_6$, and Ba$_2$CaOsO$_6$ were packed tightly in aluminum foil and loaded in identical annular cans, 3 cm in diameter. An identical empty can was also prepared for background measurements. The sample cans and empty can were sealed in a glove box containing helium gas to improve thermalization of samples at low temperatures. The cans were loaded on a three-sample carousel mounted in a closed-cycle refrigerator with a base temperature of 7 K.  Inelastic neutron scattering (INS) measurements were carried out on Ba$_2$MgOsO$_6$ and Ba$_2$ZnOsO$_6$ at a variety of temperatures above and below $T^*$, using an incident energy of E$_i$ = 60 meV while in the case of Ba$_2$CaOsO$_6$, the incident energy chosen was E$_i$ = 90 meV. The chopper settings utilized for both incident energies were the same. The chopper settings selected for the fast neutron chopper ($T_0$) and the high flux fermi chopper ($FC_1$) were $T_0$ = 240 Hz, and $FC_1$ = 120 Hz, respectively. Empty can background data sets were collected at base temperature,  and $T\,=\,200\,K$, and correspondingly subtracted from the data sets. The data sets were reduced using Mantid \cite{mantid} and analyzed using neutron scattering software DAVE \cite{dave}.

\subsection{High Intensity Neutron Powder Diffraction}

\subsubsection{Methods}

NPD data were collected on the two-axis high intensity diffractometer D20 (Institut Laue Langevin, Grenoble, France)\cite{D20} using a wavelength of $\lambda$ = 2.41 \AA\,selected by a Graphite monochromator using a take-off angle of 42$\degree$. The powder samples of Ba$_2$CaOsO$_6$ and Ba$_2$MgOsO$_6$ were those utilized for the INS measurements while a new sample of Ba$_2$ZnOsO$_6$ was synthesized for the NPD measurements. Measurements were taken at a base temperature of 10 K for 10 hrs on Ba$_2$ZnOsO$_6$ for 11 hrs on Ba$_2$MgOsO$_6$ and at 1.5 K for 11 hrs on Ba$_2$CaOsO$_6$. A higher base temperature was selected for the former compounds to avoid probing any features which could arise from the ferromagnetic ordering of the impurity phase Ba$_{11}$Os$_4$O$_{24}$ present at low levels in these compounds. High temperature data sets with the same statistics were taken at T = 50 K for Ba$_2$ZnOsO$_6$ and at T = 70 K for both Ba$_2$MgOsO$_6$ and Ba$_2$CaOsO$_6$. Data analysis was performed using the Rietveld refinement program FULLPROF \cite{Fullprof}.

\subsubsection{Magnetic Neutron Profile Calculations}

The NPD measurements performed at D20 show no evidence for magnetic order of the Os$^{6+}$ ions in Ba$_2$CaOsO$_6$, Ba$_2$MgOsO$_6$ and Ba$_2$ZnOsO$_6$. A limit for the upper bound on the magnetic moment potentially present at low temperatures in these materials can still be performed, and used for comparison with other double perovskites. This involves assuming a magnetic structure, and we employed a simple type 1 AF structure, as is observed in a variety of d$^3$ double perovskites.  FULLPROF was utilized to calculate the diffraction pattern due to the assumed magnetic structure alone, for each compound. Each calculation was performed utilizing the lattice parameters determined from fits to the low temperature data sets for each compound.  The magnetic form factor for Os$^{6+}$ was also included in this calculation, using parameters extracted from previous work done by Kobayashi \textit{et al}\cite{Kobayashi2011}. The result from this calculation was then compared with the plots showing the difference between the high temperature data set from the low temperature data set. This calculation produced a strong magnetic Bragg peak near $\sim$ 0.76 \AA$^{-1}$, corresponding to the type I AF structure, whose intensity depended on the size of the ordered moment in the structure.  The ordered moment was then adjusted until this magnetic peak was just observable above background, thus setting the upper bound to the size of a possible ordered magnetic moment in these materials.  Upper bounds for the ordered Os$^{6+}$ moments of 0.13$\mu_B$, 0.11$\mu_B$, and 0.06$\mu_B$ were obtained for Ba$_2$CaOsO$_6$, Ba$_2$MgOsO$_6$ and Ba$_2$ZnOsO$_6$, respectively.

\subsection{High Angular Resolution X-ray Diffraction}

\noindent \textit{Methods} -- Synchrotron experiments were performed on the powder diffraction station of the BL04-MSPD beamline \cite{ALBA} of the ALBA Synchrotron Light facility (Barcelona, Spain). Data were collected in transmission mode using the 13 channel multi-analyzer (MAD) setup (220 Bragg reflection of silicon crystal) which offers the highest possible instrumental angular resolution ($\Delta2\theta$ = 0.005$\degree$). In order to minimize absorption, the patterns were recorded at a short wavelength, $\lambda$ = 0.31713 \AA, with the sample filled in a 0.5 $mm$ diameter borosilicate capillary. Data were recorded in the temperature range between 20 K and 250 K using the liquid Helium flow cryostat Dynaflow \cite{cryo}.

\section{Theory}

\subsection{Single-site model, multipole moments}

We begin with an effective $J=2$ moment and incorporate the impact of a residual octahedral crystal field via
\bea
H_{\rm CEF'} = - V_{\rm eff} ({\cal O}_{40} + 5  {\cal O}_{44}),
\eea 
where the Steven's operators
\bea
{\cal O}_{40} &=& 35 J^4_z - (30 J (J+1)-25) J_z^2 + 3 J^2(J+1)^2 \nonumber \\
&-& 6 J (J+1) \\
{\cal O}_{44} &=& \frac{1}{2} (J_+^4  + J_-^4),
\eea
For $V_{\rm eff} > 0$, we find that the $J\!=\!2$ multiplet splits into a ground doublet and an excited
triplet with a gap $\Delta=120 V_{\rm eff}$. Choosing $V_{\rm eff} = 0.2$\,meV, for example, leads to a doublet-triplet
gap $\Delta \sim\! 24$\,meV.

The ground state wavefunctions in the $|m\rangle$ basis are
\bea
|\psi_{g,\uparrow}\rangle &=& |0\rangle \\
|\psi_{g,\downarrow}\rangle &=& \frac{1}{\sqrt{2}} (|2\rangle + | -2 \rangle)
\eea
while the excited state wavefunctions are given by
\bea
|\psi_{e,\pm}\rangle &=& |\pm 1\rangle\\
|\psi_{e,0}\rangle &=&  \frac{1}{\sqrt{2}} (|2\rangle - | -2 \rangle)
\eea

This sequence of level splittings for the 5d$^2$ configuration is schematically described in Fig. 1.  The rightmost level scheme results from the familiar splitting of the $5d$ 
levels in an octahedral crystal electric field (CEF).  Strong spin orbit coupling (SOC) then splits the t$_{2g}$ levels into an higher energy $j=1/2$ doublet and a lower energy quartet, 
corresponding to $j=3/2$.  The two d electrons then occupy the $j=3/2$ levels, and combine to give a $J_{\rm eff}=2$ degree of freedom.  A residual crystal field,
arising from electron-interaction induced change in the CEF splitting (denoted as CEF') gives the final non-Kramers doublet ground state with octupolar and quadrupolar
moments (discussed below), and an excited state triplet, some 15 to 30 meV above the octupolar doublet.

\begin{figure}[htbp!]
\centering
\includegraphics[width=\columnwidth]{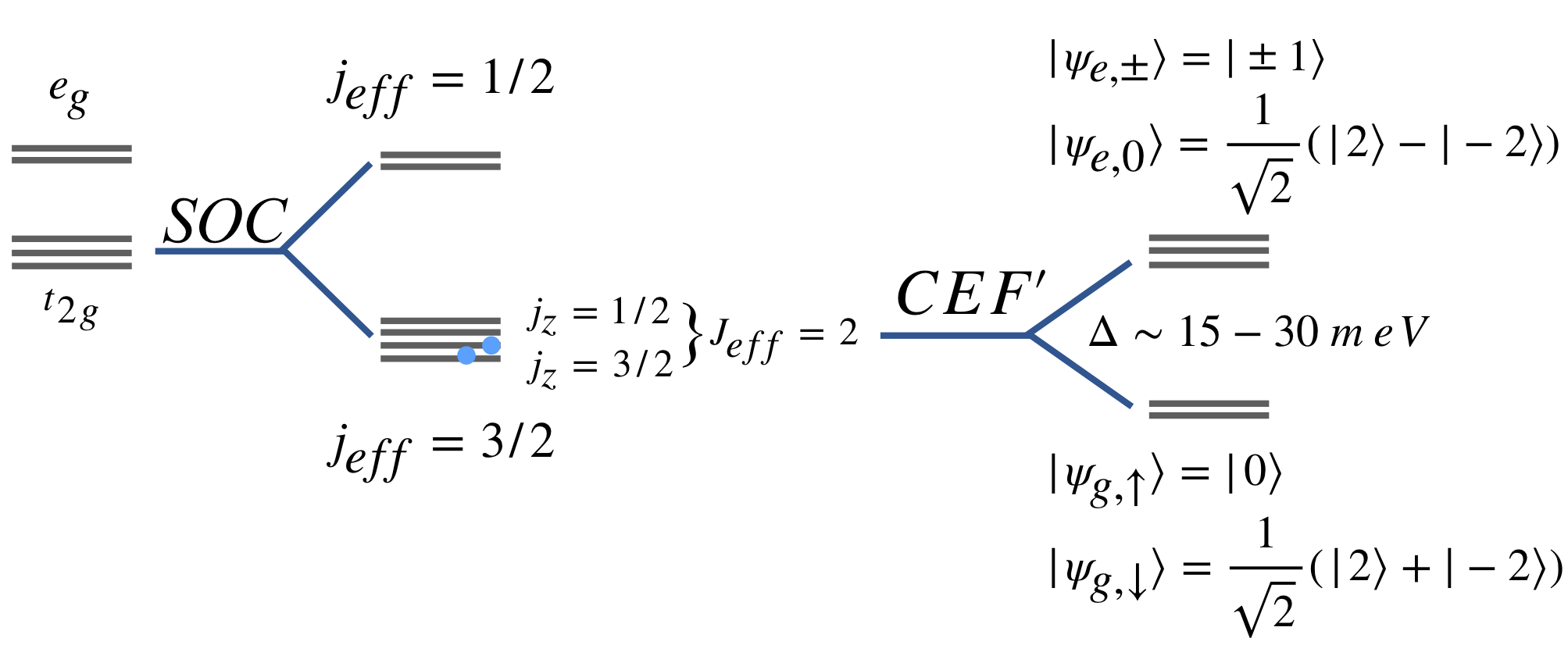}
\caption{\label{fig: Levels} The splitting of the energy levels of the $5d^2$ Os$^{6+}$ ion is schematically  illustrated. The $5d$ levels are acted on by the octahedral crystal-electric field (CEF) of the lattice, strong spin-orbit coupling (SOC), and finally a residual cubic CEF due to electron-interaction induced change in the CEF splitting (denoted here as CEF')
that results in a non-Kramers doublet ground state with quadrupolar and octupolar moments, and a weakly split off triplet of excited states some $15$-$30$ meV higher in energy. 
The ground state and excited state wavefunctions of the $d^2$ ion are explicity stated in the figure.}
\end{figure}

It is clear that the ground state manifold has vanishing matrix elements for the dipolar vector operators $\vec J$. However,
$\vec J$ can induce transitions between the ground doublet and the excited triplet. We thus attribute the observed spin gap to
this doublet-triplet gap $\Delta$.

Defining the pseudospin-$1/2$ operators $\vec\tau$ within the ground state doublet, we find that the quadrupolar operators
$(J_x^2-J_y^2) \equiv 2\sqrt{3} \tau_x$, $(3 J_z^2 - J^2) \equiv - 6 \tau_z$, while the octupolar operator 
$\overline{J_x J_y J_z} \equiv - \sqrt{3}\tau_y$ where the overline denotes symmetrization. 

\begin{figure}[t]
\centering
\begin{overpic}[width=0.3\textwidth]{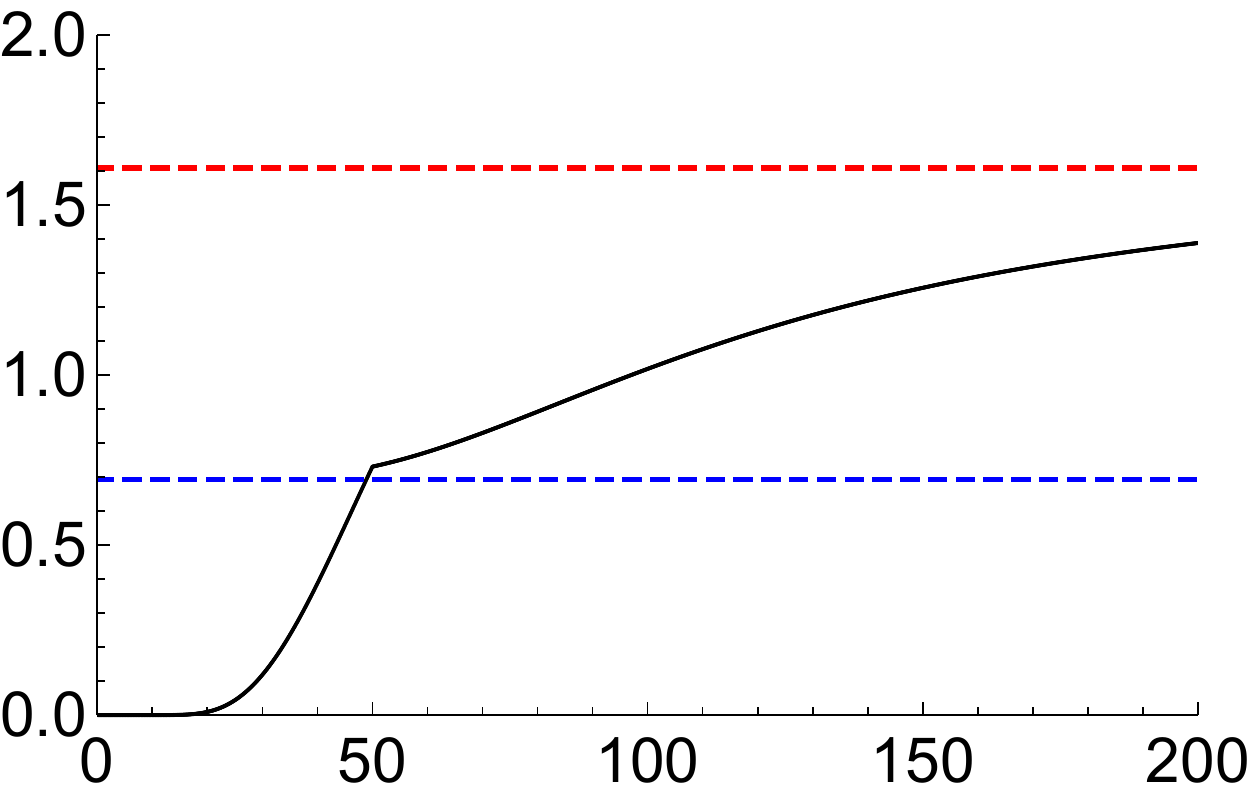}\put(50,-5){$T$ (K)}\put(-10,5){\rotatebox{90}{Entropy (units of $R$)}} \end{overpic}
\caption{Entropy of the single-site model (solid) line in units of $R$ as a function of temperature, showing the quenching of the entropy from $R \ln (5)$ at high
temperature to $\ll R \ln (5)$ (dashed lines denote $R \ln(5)$ and $R \ln(2)$) for $T  \sim 100$\,K, which is further quenched by octupolar ordering
at $T^*=50$\,K.}
\label{fig:entropy} 
\end{figure}

\begin{figure}[b]
\centering
\begin{overpic}[width=0.3\textwidth]{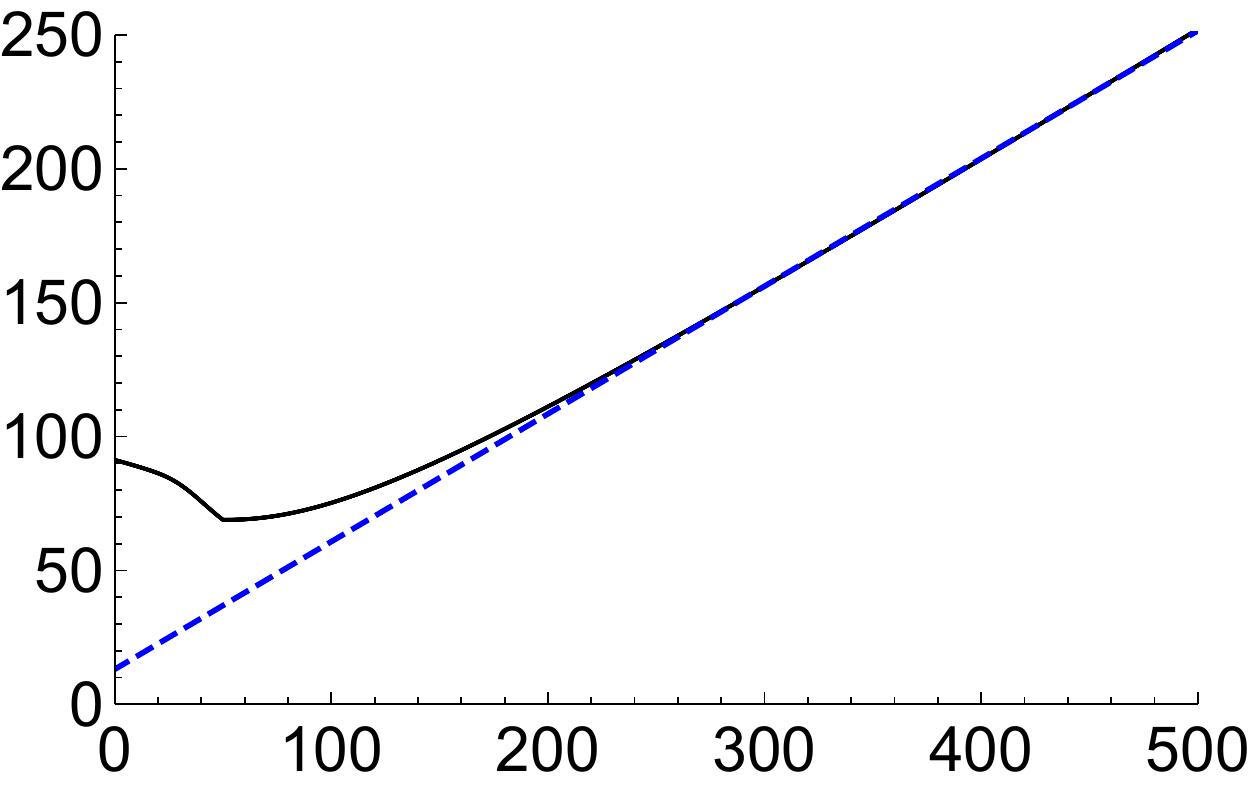}\put(50,-5){$T$ (K)}\put(-15,30){\rotatebox{90}{$1/\chi$}} \end{overpic}
\caption{Plot of the inverse susceptibility $1/\chi$ of the single-site model as a function of temperature, showing its apparent `Curie-Weiss' like
behavior at high temperature (indicated by the dashed fit) and its eventual saturation at low temperature, arising from the spin-gap, with the
kink denoting the octupolar transition.}
\label{fig:chi} 
\end{figure}

In the single-site limit, we can easily compute the following results for the
entropy and the magnetic susceptibility. Here, we pick $V_0=0.2$\,meV as above, and include a local octupolar
symmetry breaking field, in the spirit of mean field theory, via
\bea
H = H_{\rm CEF} - \phi(T) ~ \overline{J_x J_y J_z}.
\eea
As an illustrative example, we choose the symmetry breaking field $\phi(T)$ to be given by the simple form $\phi(T < T^*) = T^* \sqrt{1-T/T^*}$ and $\phi(T > T^*)=0$.

The entropy, shown in Fig.~\ref{fig:entropy}, starts at $R \ln (5)$ at high temperature, before decreasing to $\ll R \ln(5)$ for $T \lesssim 100$\,K. This residual
low temperature entropy of near $R \ln(2)$ is quenched by ordering of the doublet states induced by intersite interactions for $T < T^*$.

The inverse spin susceptibility, in units where a free spin-$S$ leads to the Curie susceptibility $S(S+1)/3T$, is shown in
Fig.~\ref{fig:chi}. It is clear that the spin gap induces an effective `antiferromagnetic Curie-Weiss' type behavior, with an intercept at
$T \!\approx\! - 30$\,K. The larger Curie-Weiss temperature $\Theta_{\rm CW} \! \approx \! -150$\,K inferred from experiments must thus stem from residual intersite interactions.

\bibliography{octupolar}